\documentclass[prl,nopacs,twocolumn,aps]{revtex4}

\usepackage{amsmath}
\usepackage{bm}
\usepackage{epsfig}
\usepackage{amssymb}

\usepackage{graphicx}
\usepackage{epsfig}
\usepackage{psfrag}

\usepackage{cancel}
\usepackage{color}

\begin{document}

\title{Facilitated spin models of dissipative quantum glasses}

\author{Beatriz Olmos}

\author{Igor Lesanovsky}

\author{Juan P. Garrahan}

\affiliation{School of Physics and Astronomy, University of
Nottingham, Nottingham, NG7 2RD, UK}

\begin{abstract}
We introduce a class of dissipative quantum spin models with local interactions and without quenched disorder that show glassy behaviour. These models are the quantum analogs of the classical facilitated spin models. Just like their classical counterparts, quantum facilitated models display complex glassy dynamics despite the fact that their stationary state is essentially trivial. In these systems, dynamical arrest is a consequence of kinetic constraints and not of static ordering. These models display a quantum version of dynamic heterogeneity: the dynamics towards relaxation is spatially correlated despite the absence of static correlations. Associated dynamical fluctuation phenomena such as decoupling of timescales is also observed. Moreover, we find that close to the classical limit quantum fluctuations can enhance glassiness, as recently reported for quantum liquids.
\end{abstract}


\maketitle

\noindent
{\bf Introduction.} A central problem in condensed-matter science is that of the glass transition. Many body systems with excluded volume interactions, such as molecular liquids, experience pronounced dynamical slowdown at high densities and/or low temperatures, to the extent that they eventually cease to relax and form the amorphous solid we call glass. The glass transition as observed experimentally is not a phase transition but a very rapid kinetic arrest. At low enough temperature or high enough density glass formers relax too slowly to be observed experimentally in equilibrium and thus behave as (non-equilibrium) solids. This solidification occurs in the absence of any evident structural ordering, in contrast to more conventional condensed matter systems: in glass formers thermodynamics changes apparently very little but dynamics changes dramatically. Dynamic arrest like that of glasses is a generic phenomenon in condensed matter; for recent reviews see \cite{Cavagna09,Chandler10,Berthier11,Binder11}.

While the first hallmark of glass formers is kinetic arrest, the second is {\em dynamical heterogeneity} \cite{DH}. Glass formers appear structurally homogeneous, but their dynamics is highly heterogeneous: as they slow down spatial dynamical correlations emerge and these become more pronounced the longer the relaxation times. One theoretical perspective on glasses in which dynamical heterogeneity appears naturally is that of {\em dynamical facilitation}, which posits that the origin of glassy slowing down is not to be found in thermodynamic ordering \cite{thermo} but in effective constraints in the dynamics (see \cite{Chandler10} for a review). From this perspective, slowdown, heterogeneity and other fluctuation features of glasses are rooted in the complex structure of trajectory space. This theory has emerged from the study of a class of idealized lattice systems, so-called kinetically constrained models, of which the simplest representatives are the facilitated spin models \cite{Ritort03}.

\begin{figure}[th]
\includegraphics[width=.85\columnwidth]{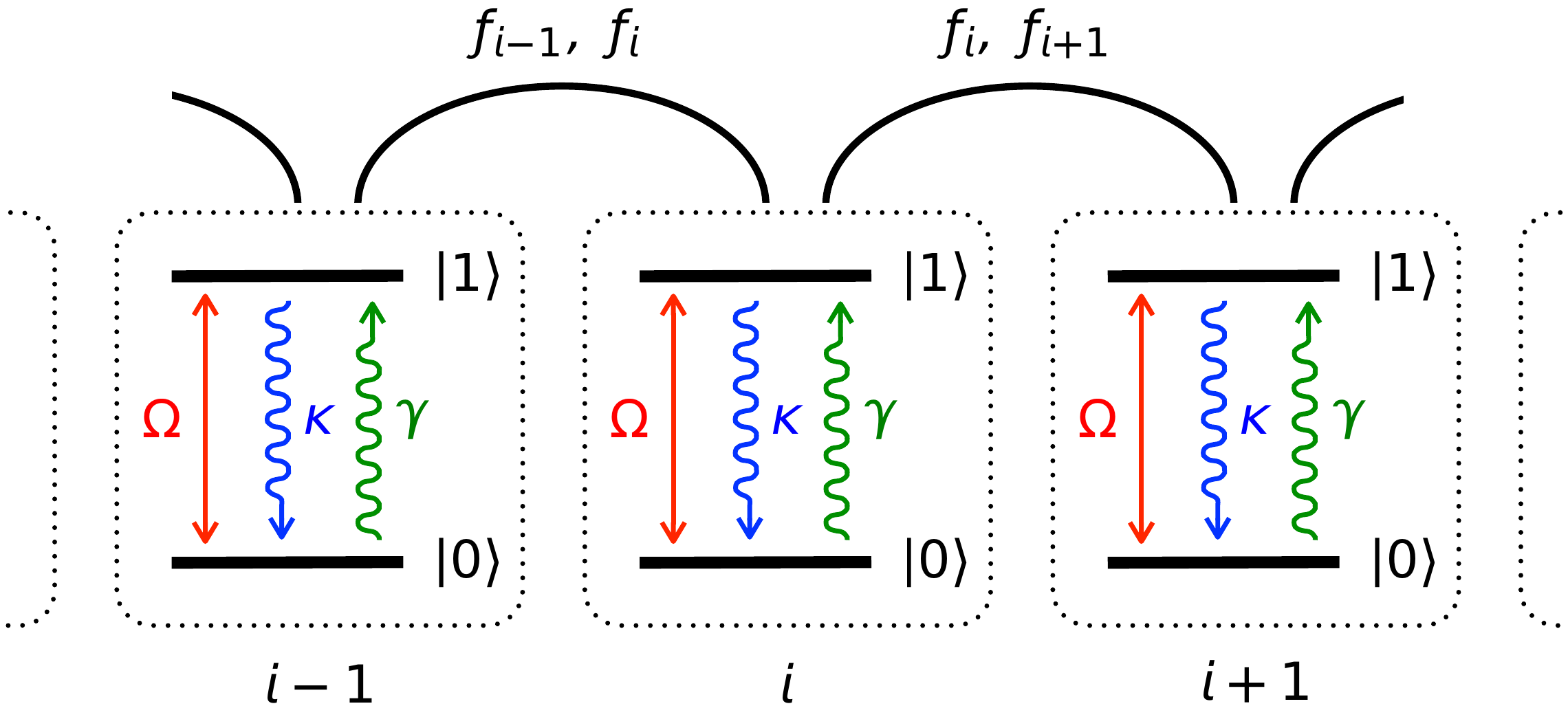}
\caption{{\em Quantum facilitated models.} A lattice of two-level systems with coherent and incoherent excitation/de-excitation.  Interaction is via kinetic constraints $f_i$ which make the rates on site $i$ dependent on the state of its neighbours.  The stationary state is trivial and non-interacting, while the dynamics is highly correlated and glassy when $\kappa \gg \Omega,\gamma$.
}
\label{model}
\end{figure}

Quantum glasses, just like their classical counterparts, are of much current interest, among other reasons due to their relevance to issues like supersolidity \cite{Biroli08}, quantum annealing \cite{Das08}, glassiness in electronic systems \cite{Amir09}, thermalization and many-body localization \cite{Pal10}, and arrest in quantum fluids \cite{Markland11}.  Central questions in quantum many-body systems which display glassy behaviour are understanding the interplay between static and dynamic properties, the relevance of quantum versus classical fluctuations, and the emergence of spatial correlations in the relaxation dynamics. In this work we take a first step towards addressing these issues from a dynamical facilitation perspective by introducing and studying a class of open quantum lattice systems that are the quantum analogs of classical facilitated spin models of glasses. These quantum spin systems, which are free of quenched disorder and have local interactions, display complex glassy dynamics despite the fact that their static properties are trivial. They also show a quantum version of dynamic heterogeneity, a feature we expect to be central to the dynamics of quantum glasses in general.

\noindent
{\bf Quantum facilitated models.} Our aim is to construct a class of strongly interacting dissipative quantum many-body systems with the following two properties: firstly, interactions do not play an essential role in the statics and, as a consequence, the steady state of the system has the form of a direct product of single-site density matrices, like in a non-interacting problem. Secondly, the interactions make the dynamics and, in particular, the relaxation to the steady state, highly spatially correlated.

Let us consider the lattice system of Figure \ref{model}. On each lattice site $i=1,\ldots,N$ there is a quantum two-level system, and we identify the two levels with basis states $|0\rangle_i$ and $|1\rangle_i$. These two levels are coupled coherently via an on-site Hamiltonian $h_i = \Omega \sigma_i^x$,
where $\Omega$ is the coherent coupling strength and $\sigma_i^x \equiv |0 \rangle_i \langle 1| + |1\rangle_i \langle 0|$. Each two-level system also interacts with a thermal bath that can cause incoherent excitation and de-excitation with rates $\gamma$ and $\kappa$, respectively. In the absence of interaction between the sites, and under the standard Markovian approximation for the thermal bath \cite{Gardiner04}, the open quantum dynamics of each site is described by the master equation
\begin{eqnarray}
\dot{\rho}_i ={\cal L}_i(\rho_i) &\equiv& -i \left[ h_i , \rho_i\right]
+\kappa \sigma^-_i \rho_i \sigma^+_i
- \frac{\kappa}{2} \left\{  \sigma^+_i \sigma^-_i , \rho_i \right\}
\nonumber \\
&&+\gamma \sigma^+_i \rho_i \sigma^-_i
- \frac{\gamma}{2} \left\{ \sigma^-_i \sigma^+_i , \rho_i \right\},
\label{mastereqi}
\end{eqnarray}
with $\sigma_i^\pm$ being the ladder operators defined in terms of the usual Pauli matrices as $\sigma_i^\pm=\left(\sigma^x_i\pm i\sigma^y_i\right)/\sqrt{2}$. This equation has as stationary solution $\rho_i^{({\rm st})} \equiv \lambda_u |u\rangle_i \langle u| + \lambda_e |e\rangle_i \langle e|$, where the \emph{unexcited} ($|u\rangle_i$) and \emph{excited} ($|e\rangle_i$) states can be written as a superposition of $|0\rangle_i$ and $|1\rangle_i$ and $\lambda_{u,e}$ are their corresponding stationary occupation probabilities \cite{SM}. When the two-level systems are non-interacting the stationary density matrix for the whole system is simply the direct product
\begin{equation}
\varrho^{({\rm st})}_{\rm ni} = \bigotimes_{i=1}^N \rho_i^{({\rm st})}.
\label{rhost}
\end{equation}

We now convert the system into an interacting problem with the the same trivial stationary state. We introduce interactions between nearest neighbouring sites in the following manner: we condition the rates for both coherent and incoherent changes at the $i$-th site to the state of neighboring sites, i.e., $\Omega \to f_i \Omega$, $\kappa \to f_i \kappa$ and $\gamma \to f_i \gamma$, where $f_i$ is a projection operator that acts only on the nearest neighbours of $i$, see Fig.\ \ref{model}. The many-body master equation now reads
\begin{eqnarray}
\dot{\varrho} = {\mathbb W}(\varrho) \equiv -i\left[H,\varrho\right] + \sum_{i=1}^N
\left(
L_i \varrho L_i^\dag -\frac{1}{2} \left\{ L_i^\dag L_i, \varrho \right\}
\right. \nonumber \\
\left.
J_i \varrho J_i^\dag -\frac{1}{2} \left\{ J_i^\dag J_i, \varrho \right\}
\right)
\label{mastereq}
\end{eqnarray}
with Hamiltonian and Lindblad \cite{Gardiner04} operators given by
\begin{equation}
H = \sum_{i=1}^N \Omega f_i \sigma_i^x , \;\;\;
L_i = \sqrt{\kappa} f_i \sigma_i^- , \;\;\;
J_i = \sqrt{\gamma} f_i \sigma_i^+ .
\label{LJ}
\end{equation}
The operators $f_i$ represent {\em kinetic constraints}.  Under certain general conditions for $f_i$, which we derive in the Supplemental Material \cite{SM}, the stationary state of the interacting problem is the trivial one (\ref{rhost}), i.e. $\varrho^{({\rm st})} = \varrho^{({\rm st})}_{\rm ni}$. Here we focus on two specific choices for the constraints which define the two models we study in detail (in dimension one, as higher dimensional generalisations are straightforward):
\begin{eqnarray}
{\rm qFA:} &&
f_i \equiv P^{(e)}_{i+1} + P^{(e)}_{i-1} - P^{(e)}_{i+1} P^{(e)}_{i-1}
\label{qFA}
\\
{\rm qEast:} &&
f_i \equiv P^{(e)}_{i+1},
\label{qEast}
\end{eqnarray}
with $P^{(e)}_i \equiv |e\rangle_i \langle e|$ being projectors on the excited state on the $i$-th site.

\begin{figure}
\includegraphics[width=.75\columnwidth]{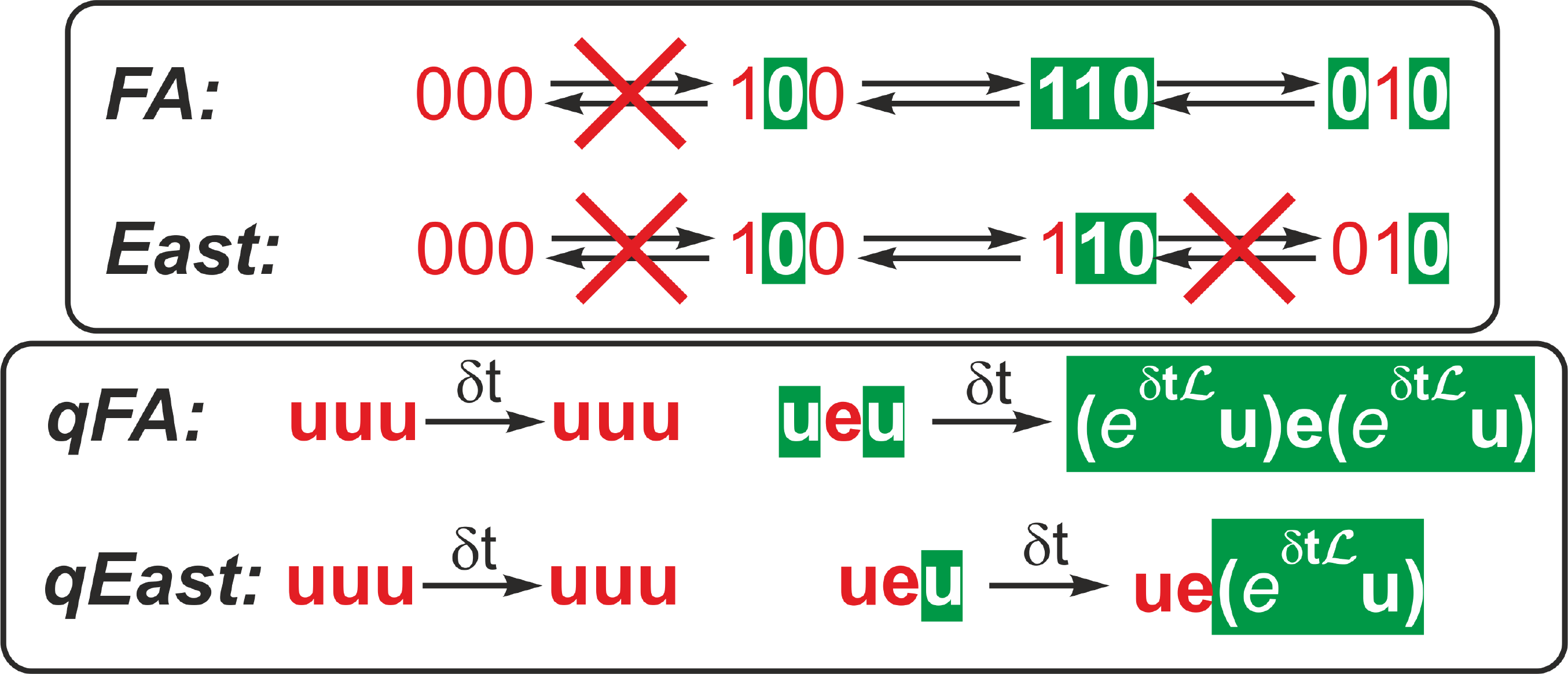}
\caption{{\em Kinetic constraints.} For the classical models the upper panel shows three sites in a configuration, assuming all other sites are in the $0$ state. For the quantum models the bottom panel shows three sites of a product basis state for the density matrix, where we use the notation ${\boldsymbol u} \equiv |u\rangle \langle u|$ and ${\boldsymbol e} \equiv |e\rangle \langle e|$. Red (dark on white) symbols represent sites that cannot evolve.  Green (white on dark) symbols represent sites where the state of neighboring sites allow for their evolution.
}
\label{fig:Diagram}
\end{figure}

The lattice models defined by Eqs.\ (\ref{mastereq})-(\ref{LJ}) are quantum versions of the facilitated spin models for classical glasses \cite{Ritort03}. In particular, the choice of kinetic constraint (\ref{qFA}) defines a quantum Fredrickson-Andersen (qFA) model, while that of (\ref{qEast}) a quantum East (qEast) model \cite{Ritort03}. The classical Fredrickson-Andersen (FA) and East models are recovered in the limit of vanishing coherent coupling, $\Omega = 0$.  Figure \ref{fig:Diagram} sketches the effect of the kinetic constraints.  For the classical FA and East models (upper panel), a site cannot change if both neighbours are in the state 0. In the FA model an excitation can facilitate changes to its left or its right, while in the East model facilitation is directional (e.g. only to the right as shown in the figure).

In analogy to the classical case, in both the qFA and qEast models (bottom panel in Fig.\ \ref{fig:Diagram}) there can be no evolution in a site surrounded by unexcited neighbours [see Eqs.\ (\ref{qFA},\ref{qEast})]. In the qFA an excitation, like the central one in the rightmost sketch, facilitates both its neighbours which evolve with the single site master operator of Eq.\ (\ref{mastereqi}). In contrast, in the qEast the central excitation can only facilitate dynamics to its right.

\begin{figure*}[th]
\includegraphics[width=2\columnwidth]{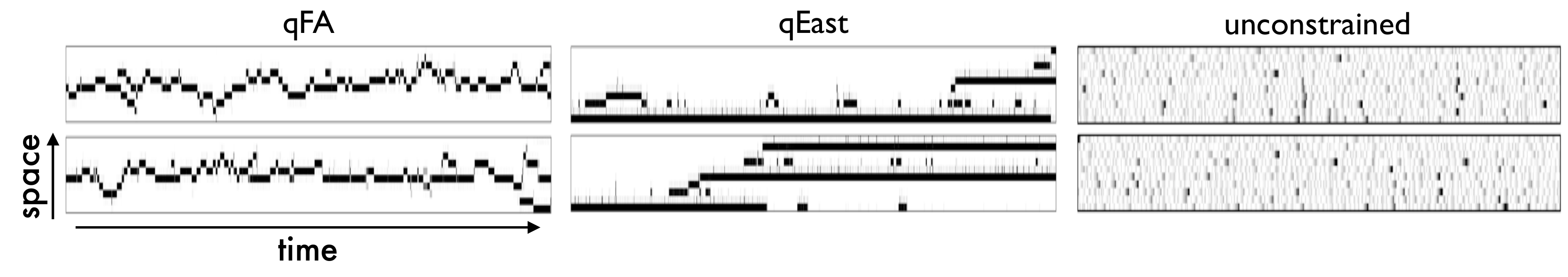}
\caption{{\em Structure in trajectory space and dynamic heterogeneity}.
Quantum trajectories of the qFA, qEast and unconstrained models, from QJMC simulations for a ring of $N=10$ spins. We plot $\langle P^{(e)}_i \rangle(t) \in [0,1] \equiv [{\rm white},{\rm black}]$ for all sites $i=1$ to $N$. The rates $(\kappa,\Omega,\gamma)$ for the trajectories shown are: $(1,0.25,0)$ for the qFA and $(1,0.4,0)$ for the qEast and the unconstrained ones. Space-time correlations are evident in the trajectories of the constrained models, and absent in the unconstrained system.}
\label{trajectories}
\end{figure*}

It is interesting to note that, in addition to the state of Eq.\ (\ref{rhost}), there is a second stationary state due to the kinetic constraints. This pure state, $\varrho^{\rm (dark)} \equiv |u \cdots u\rangle \langle u \cdots u|$, is a ``dark state" which is annihilated by the Hamiltonian and all the Lindblad operators independently, since $f_i \varrho^{\rm (dark)} = \varrho^{\rm (dark)} f_i =0$ for all $i$.  The dynamics defined by (\ref{mastereq}-\ref{LJ}) is therefore reducible: It has one irreducible and ``inactive" partition composed solely of $\varrho^{\rm (dark)}$ and a second irreducible and ``active" partition composed of everything else. In the thermodynamic limit, $N \to \infty$, the stationary state of the active partition becomes $\varrho^{\rm (st)}$ up to corrections which vanish exponentially with $N$. In what follows we only consider dynamics in this partition. Nevertheless, the existence of the inactive state has important consequences for the dynamical behaviour of the system as we discuss below.

\noindent
{\bf Dynamical heterogeneity and slowdown of the relaxation.}
We are interested in the dynamics of our quantum facilitated spin models in the regime where the coherent coupling is weak and/or the temperature is low, i.e. $\kappa \gg \Omega,\gamma$. In this regime $\lambda_u \gg \lambda_e$ (see \cite{SM}), that is, the stationary state is such that very few sites are in the excited state $|e\rangle$. This however leads to a conflict with the dynamics, as the kinetic constraint on a site vanishes unless some of its neighbours are in the excited state, and in this regime most sites will be surrounded by unexcited sites. The consequence is a pronounced slowdown and the emergence of collective dynamics.

Let us consider how excitations propagate in the quantum models we propose. Typically, at low $\Omega/\kappa,\gamma/\kappa$, excitations will be isolated. We denote a state with such an excitation at site $i$ by $\cdots {\boldsymbol u}_{i-1} {\boldsymbol e}_i {\boldsymbol u}_{i+1} \cdots$ (see Fig. \ref{fig:Diagram}). As we have explained, the $i$-th site cannot evolve, but it can facilitate evolution of its neighbours.
Let us first consider the case of the qFA model: here, the initial excitation can facilitate the excitation of either of the neighbouring sites. This new excitation subsequently facilitates the original site, which can now de-excite. The first process could be either virtual or real, so is limited by the rates $\Omega$ or $\gamma$. The second process will be dominated by incoherent de-excitation of rate $\kappa$. In the classical limit, the outcome of such sequence is that an isolated excitation hops one site. In the qFA model excitations also propagate diffusively, with an effective diffusion constant that vanishes in the limit $\Omega/\kappa,\gamma/\kappa \to 0$, in analogy to the classical FA model at low temperatures \cite{Ritort03}.
Dynamics of the qEast model is even more complex. Due to the directionality of the kinetic constraint, Eq.\ (\ref{qEast}), the last step in the hopping sequence sketched above is not allowed and the relaxation is therefore {\em hierarchical} \cite{Ritort03,Chandler10}.

Figure \ref{trajectories} shows examples of quantum trajectories from quantum jump Monte Carlo (QJMC) simulations \cite{Plenio98}. The trajectories for the qFA clearly show the diffusive nature of the dynamics: when initialized from a single site in the $|e\rangle$ state and the rest in the $|u\rangle$, this initial excitation propagates in a diffusive manner. Occasionally, excitations branch out, or coalesce. Due to the kinetic constraints excitations have to form connected chains in space and time \cite{Chandler10}, as illustrated in the figure. The hierarchical relaxation dynamics of the qEast model is also observed in the trajectories of Fig.\ \ref{trajectories}. For comparison we also show trajectories under similar conditions for the unconstrained problem [see Eq.\ (\ref{mastereqi})]. These trajectories are featureless, as there is no interaction between the sites, who then evolve independently. Note that all three systems shown in Fig.\ \ref{trajectories} possess the \emph{same stationary properties}, despite the fact that their dynamics (and the approach to such stationary state) is obviously different. It is instructive to compare the quantum trajectories of Fig.\ \ref{trajectories} to those of the corresponding classical models, see e.g. Fig.\ 2 of Ref.\ \cite{Garrahan02}.

\begin{figure}
\includegraphics[width=\columnwidth]{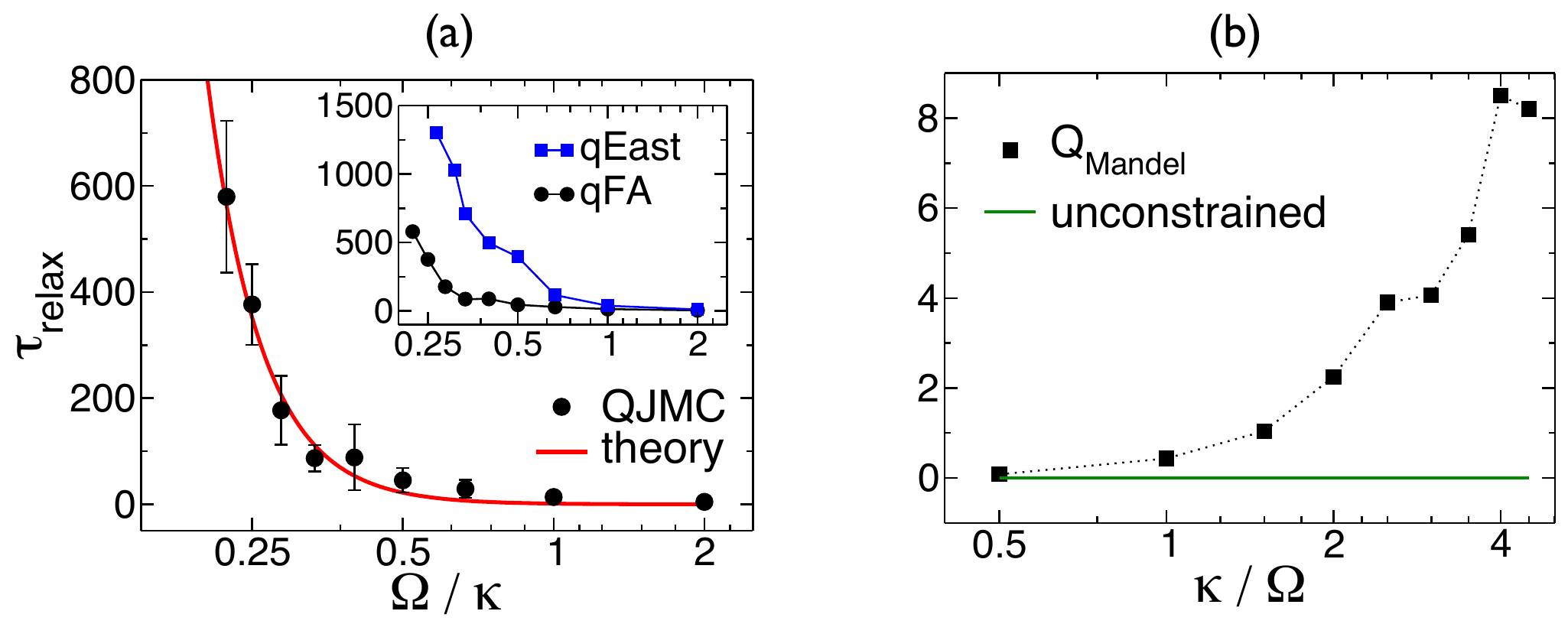}
\caption{{\em Glassy slowing down and dynamical correlations.} (a) Average relaxation time $\tau_{\rm relax}$ as a function of $\Omega/\kappa$ (at $\gamma=0$) in the qFA model. The dots are results from QJMC simulations. The line is the theoretical expectation for relaxation via activated diffusion of excitations: since the initial state has a single excitation, in a lattice of size $N$ with periodic boundary conditions we expect $\tau_{\rm relax} \approx (2/N) \sum_{l=1}^{N/2} \tau(l)$, where $\tau(l) \approx l^2/D$. Inset: $\tau_{\rm relax}$ for the qEast model. (b) Fluctuations in waiting times between quantum jump events for the same qFA system, quantified via the Mandel parameter $Q_{\rm Mandel}$. A value of $Q_{\rm Mandel} > 0$ is indicative of correlated fluctuations in the dynamics.}
\label{times}
\end{figure}

The effective diffusion constant for propagation of excitations in the qFA model can be estimated analytically when $\Omega,\gamma \ll \kappa$ by carrying out an adiabatic elimination of fast degrees of freedom, or by numerical simulation of the evolution operator (\ref{mastereq}) in small lattices (see Supplemental Material \cite{SM}). These results suggest that the diffusion constant reads
\begin{equation}
D = \left( \frac{\gamma}{2} - \frac{\gamma^2}{6\kappa} \right) - \frac{4 \Omega^2}{\kappa^2} \left( \gamma - \frac{2 \gamma^2}{\kappa} \right) + \frac{8 \Omega^4}{\kappa^3} + \cdots
\label{D}
\end{equation}
In the limit of zero temperature, $\gamma=0$, the diffusion constant is $D = 8\Omega^4/\kappa^3$ to leading order in $\Omega$. Relaxation of a site initially in an unexcited state will depend on how long it takes for the nearest excitation to diffuse to it. If the distance is $l$, then the time to relax will be $\tau(l) \approx l^2/D$. Figure \ref{times}(a) shows that this diffusive argument accounts very well for the relaxation time $\tau_{\rm relax}$ obtained from QJMC simulations in the qFA model. As expected, the relaxation slows down very markedly with the decrease of quantum fluctuations. The inset to Fig.\ \ref{times}(a) shows that the relaxation time in the qEast model grows with decreasing $\Omega$ even faster than in the qFA as a consequence of the more restrictive kinetic constraints.

From the trajectories of the qFA and qEast shown in Fig.\ \ref{trajectories} we see that, in the constrained cases considered, when the dynamics is slow it is also heterogeneous. Regions of space empty of excitations are slow to relax, as they need excitations external to the region to propagate into it for the dynamics to take place. The larger these empty regions the longer they take to evolve. These slow ``space-time bubbles" \cite{Garrahan02} make the dynamics fluctuation dominated.  Their origin can be traced back to the dark state $\varrho^{\rm (dark)}$: even if this is never accessed from the active partition, at low $\Omega/\kappa,\gamma/\kappa$ mesoscopic regions devoid of excitations look locally like the dark state; it is these spatial rare regions that give rise to dynamic heterogeneity.  Furthermore, in the limit of $N \to \infty$, dynamic bottlenecks can become so pronounced that these systems display non-equilibrium (or trajectory) phase transitions between active and inactive dynamical phases, of the kind discussed in \cite{Garrahan10}.

A manifestation of dynamic heterogeneity is in the fluctuations of the waiting times $t_w$ between quantum jump events in each site. When the dynamics is spatially correlated, quantum jumps are non-Poissonian and the corresponding distribution of waiting times is non-exponential. This can be quantified by the Mandel-Q parameter for the waiting times, $Q_{\rm Mandel} \equiv \langle t_w^2 \rangle / \langle t_w \rangle^2 - 2$ \cite{Barkai04}. For correlated (bunched) quantum jumps we have that $Q_{\rm Mandel} > 0$, indicating that the fluctuations in the waiting times can be much larger than what is expected from a Poisson process. Figure \ref{times}(b) shows that this is the case for the qFA model. In the glass literature this is often referred to as ``decoupling" \cite{DH,Chandler10}.

\begin{figure}
\includegraphics[width=.55\columnwidth]{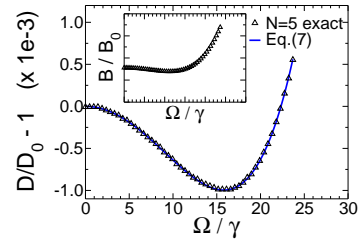}
\caption{{\em Enhanced glassiness due to weak quantum fluctuations.} Effective diffusion constant $D$ in the qFA model at $\gamma \neq 0$: It initially decreases as the strength of quantum fluctuations is increased from the classical limit ($D_0$ at $\Omega=0$). Weak quantum fluctuations hinder, rather than promote, relaxation. Inset: Similar behaviour of the branching rate $B$.}
\label{reentrance}
\end{figure}

\noindent
{\bf Interplay of quantum and classical fluctuations.} For the system to relax it is necessary to excite sites, which can be done incoherently, i.e., classically, with rate $\gamma$, or coherently with rate $\Omega$. It would then seem that adding quantum fluctuations to a classical facilitated system should always aid relaxation. But this is not the case. From Eq.\ (\ref{D}) we see that in the qFA model the first correction to the classical diffusion constant is negative, and only for larger $\Omega$ quantum contributions become positive; see Figure \ref{reentrance}. This means that weak quantum fluctuations enhance dynamical arrest by decreasing the propagation rate of excitations. The inset to Fig.\ \ref{reentrance} shows that this also holds for the rate $B$ for branching of excitations \cite{SM}. This phenomenon of enhanced glassiness due to weak quantum fluctuations near the classical limit appears to be similar to that reported recently for glass forming quantum liquids in Ref.\ \cite{Markland11}.

\noindent
{\bf Outlook.}
We expect the dynamics of glassy quantum many-body systems with local interactions to display the essential feature of the idealized models we introduced in this paper, namely a spatially fluctuating and heterogeneous dynamics which cannot be simply inferred from  static behaviour.  To understand this correlated dynamics one would need to uncover the effective kinetic constraints.  In this sense, we envisage quantum facilitated models as an effective description of the more complex glassy many-body quantum dynamics.
Given the recent progress in quantum many-body physics with cold atoms it may be possible to directly implement and study constrained models in experiments.   For example, Rydberg atoms \cite{Gallagher84} exhibit strong interactions that have been shown to lead to constraints in the coherent dynamics
\cite{Lesanovsky11,Ji11,Ates12} which, in conjunction with atomic decay, can give rise to pattern formation in quantum jump trajectories \cite{Lee12}.  Furthermore, recent work focusing on the
engineering of system-bath coupling \cite{Diehl08,Diehl10-2} outlines a path for devising the constrained many-body jump operators required by the models presented here.

\begin{acknowledgments}
This work was funded in part by EPSRC Grant no. EP/I017828/1 and Leverhulme Trust grant no. F/00114/BG. B.O. also acknowledges funding by Fundaci\'on Ram\'on Areces.
\end{acknowledgments}


\appendix* \section*{SUPPLEMENTAL MATERIAL}

\noindent{\bf Non-interacting problem.}
Here we obtain the stationary density matrix $\rho^{(\mathrm{st})}_{i}$ corresponding to Eq.\ (\ref{mastereqi}). It reads
\begin{equation*}
  \rho^{(\mathrm{st})}_{i}=\frac{1}{8\Omega^2+\left(\kappa+\gamma\right)^2} \left(\begin{array}{cc}
    4\Omega^2+\gamma(\kappa+\gamma) & -2i\Omega(\kappa-\gamma)\\
    2i\Omega(\kappa-\gamma) & 4\Omega^2+\kappa(\kappa+\gamma)
  \end{array}\right),
\end{equation*}
and can be diagonalized such that $\rho_i^{({\rm st})} \equiv \lambda_u |u\rangle_i \langle u| + \lambda_e |e\rangle_i \langle e|$. For $\kappa\gg\gamma,\Omega$, and to leading order in $\Omega/\kappa,\gamma/\kappa$ one obtains that
\begin{equation*}
\lambda_u \approx 1 - \frac{\gamma}{\kappa} - \frac{16\Omega^4}{\kappa^4}
, \;\;\;
\lambda_e \approx \frac{\gamma}{\kappa} + \frac{16\Omega^4}{\kappa^4} ,
\end{equation*}
and
\begin{eqnarray*}
  \left|u\right>\approx\left(\begin{array}{c}
    -\frac{2i\Omega(\kappa-\gamma)}{\kappa^2}\\
    1-\frac{2\Omega^2}{\kappa^2}
  \end{array}\right),\qquad
  \left|e\right>\approx\left(\begin{array}{c}
    i-i\frac{2\Omega^2}{\kappa^2}\\
    \frac{2\Omega(\kappa-\gamma)}{\kappa^2}
  \end{array}\right).
\end{eqnarray*}
Note that, in the limit of $\Omega\rightarrow0$, the two states $\left|u\right>$ and $\left|e\right>$ can be identified with $\left|0\right>$ and $\left|1\right>$, respectively.

\bigskip
\noindent{\bf Kinetic constraints.}
We are interested in introducing a constraint in the model such that the stationary state solution is the same as in the unconstrained case. This can be done by means of the transformation
\begin{equation*}
H = \sum_{i=1}^N \Omega f_i \sigma_i^xf_i^\dag , \;\;\;
L_i = \sqrt{\kappa} f_i \sigma_i^- , \;\;\;
J_i = \sqrt{\gamma} f_i \sigma_i^+ .
\end{equation*}
where $f_i$ represents the kinetic constraint. The new master equation can be written as
\begin{eqnarray*}
  \dot\varrho&=&\sum_{i=1}^N\left(f_i{\cal L}_i(\varrho)f_i^\dag+A_i(\varrho)\right),
\end{eqnarray*}
with
\begin{eqnarray*}
  A_i(\varrho)&=&-i\Omega\left(f_i\sigma_i^x\left[f_i^\dag,\varrho\right]+\left[f_i,\varrho\right]\sigma_i^xf_i^\dag\right)\\
&&-\frac{\kappa}{2}\left(\left\{\sigma_i^+f_i^\dag f_i\sigma_i^-,\varrho\right\}-f_i\left\{\sigma_i^+\sigma_i^-,\varrho\right\}f_i^\dag\right)\\
&&-\frac{\gamma}{2}\left(\left\{\sigma_i^-f_i^\dag f_i\sigma_i^+,\varrho\right\}-f_i\left\{\sigma_i^-\sigma_i^+,\varrho\right\}f_i^\dag\right).
\end{eqnarray*}
For this master equation to posses the same stationary solution as the unconstrained problem, we need that $A_i(\varrho^{(\mathrm{st})}_\mathrm{ni})=0$, which is accomplished only if the following three properties of the constraints are fulfilled: (i) they are projectors, $f_i = f_i^\dag = f_i^2$ or, more generally, normal operators $f_i f_i^\dag = f_i^\dag f_i$; (ii) the constraint at site $i$ commutes with all operators at that site, $[f_i , \sigma_i^x] = [f_i , \sigma_i^y] = [f_i , \sigma_i^z] = 0$; (iii) all constraints commute with the non-interacting stationary density matrix, $[f_i , \varrho^{({\rm st})}_{\rm ni}]=0 ~ \forall i$. The constraints of Eqs.\ (\ref{qFA},\ref{qEast}) which define the qFA and qEast models satisfy these properties.

Condition (iii) is not present in classical models; it is required as quantum density matrices in general are non-diagonal. It strongly restricts the class of operators that can be taken as kinetic constraints. In particular, on each site they have to be diagonal in the $\left\{ |u\rangle,|e\rangle \right\}$ basis. Equations (\ref{mastereq}-\ref{LJ}) and conditions (i-iii) define the class of {\em quantum facilitated spin models}, of which the qFA and qEast are two representatives.

\bigskip
\noindent
{\bf Diffusion in the qFA model.}
Here we will obtain an analytic approximation for the diffusion constant in the qFA model in the regime where $\kappa\gg\Omega,\gamma$ using the method of the adiabatic elimination of fast variables \cite{Gardiner85}.

Let us consider the case of a system formed only by two sites, so that the basis is $\left\{\left|uu\right>,\left|eu\right>,\left|ue\right>,\left|ee\right>\right\}$. As discussed in the main text, the inactive state $\left|uu\right>$ is dynamically disconnected from any of the other states of the basis. Let us assume that initially no population is in this state, so that one can look at the dynamics of the reduced density matrix spanned only by the states $\left\{\left|ee\right>,\left|eu\right>,\left|ue\right>\right\}$. The master equation of the system is given by (\ref{mastereq}), and we divide it into two parts,
\begin{equation*}
  \dot{\varrho}=\left[{\mathbb W}_\kappa+{\mathbb W}_0\right](\varrho),
\end{equation*}
with the first term being proportional to the decay rate $\kappa$ which we assume is much larger than $\gamma,\Omega$ and the second term being, at most, proportional to $\gamma$ and $\Omega$. In particular, one can see that the action of the first operator on the density matrix is given by
\begin{equation*}
  {\mathbb W}_{\kappa}(\varrho)=\left(\begin{array}{ccc}
    -2\kappa\varrho_{ee,ee} & -\kappa\varrho_{ee,eu} & -\kappa\varrho_{ee,ue}\\
    -\kappa\varrho_{eu,ee} & \kappa\varrho_{eu,eu} & 0\\
    -\kappa\varrho_{ue,ee} & 0 & \kappa\varrho_{ue,ue}
  \end{array}\right).
\end{equation*}

We now proceed with the adiabatic elimination of fast variables, i.e., we look at the time evolution of the systemat typical time scales much larger than $\kappa^{-1}$. First, we introduce the projection operator ${\cal P}=\lim_{t\rightarrow\infty} e^{{\mathbb W}_\kappa t}$ that projects on the null space of ${\mathbb W}_\kappa$, such that $\alpha \equiv {\cal P}\varrho$ is the stationary solution of $\dot{\alpha}={\mathbb W}_\kappa(\alpha)=0$. From the original master equation, and assuming that initially the state $\left|ee\right>$ is not populated, the problem can be reduced to the effective time-evolution of the $2\times2$ density matrix $\alpha$ expressed in the basis $\left\{\left|eu\right>,\left|ue\right>\right\}$. The time-evolution of $\alpha$ is given by $\dot{\alpha}={\mathbb W}_\alpha \alpha$ with
\begin{equation}\label{Eq:alpha}
  {\mathbb W}_\alpha={\cal P}{\mathbb W}_0-{\cal P}{\mathbb W}_0\left[{\cal Q}{\mathbb W}_0+{\mathbb W}_\kappa\right]^{-1}{\cal Q}{\mathbb W}_0,
\end{equation}
where ${\cal Q}=1-{\cal P}$.
We can obtain from here, at least formally, the exact solution of the $N=2$ problem which, in the the large $\kappa$ limit yields
\begin{equation*}
  \dot\alpha\approx\left[{\cal O}_1+{\cal O}_2+{\cal O}_3+{\cal O}_4\right]\alpha
\end{equation*}
with the operators ${\cal O}_k$ representing the contribution of order $k$ in $\Omega$ and $\gamma$: ${\cal O}_1={\cal P}{\mathbb W}_0$, ${\cal O}_2=A$, ${\cal O}_3=B$ and ${\cal O}_4=C+EA$, with
\begin{eqnarray*}
  A&=&-{\cal P}{\mathbb W}_0{\mathbb W}_\kappa^{-1}{\cal Q}{\mathbb W}_0\\
  B&=&{\cal P}{\mathbb W}_0{\mathbb W}_\kappa^{-1}{\cal Q}{\mathbb W}_0{\mathbb W}_\kappa^{-1}{\cal Q}{\mathbb W}_0\\
  C&=&-{\cal P}{\mathbb W}_0{\mathbb W}_\kappa^{-1}{\cal Q}{\mathbb W}_0{\mathbb W}_\kappa^{-1}{\cal Q}{\mathbb W}_0{\mathbb W}_\kappa^{-1}{\cal Q}{\mathbb W}_0\\
  E&=&-{\cal P}{\mathbb W}_0{\mathbb W}_\kappa^{-2}{\cal Q}{\mathbb W}_0.
\end{eqnarray*}
Up to fourth order in $\Omega$ and second in $\gamma$, the effective equation of motion is given by
$\dot{\alpha}={\mathbb W}_\mathrm{eff}\alpha$ with the effective master operator written in matrix form as
\begin{equation*}
  {\mathbb W}_\mathrm{eff}=\left(\begin{array}{cccc}
    -{\cal D} & 0 & 0 & {\cal D}\\
    0 & {\cal C} & 0 & 0\\
    0 & 0 & {\cal C} & 0\\
    {\cal D} & 0 & 0 & -{\cal D}
  \end{array}\right)
\end{equation*}
where
\begin{eqnarray*}
  {\cal D} &=& \frac{\gamma}{2} - \frac{4 \Omega^2}{\kappa^2} \left( \gamma - \frac{2 \gamma^2}{\kappa} \right) + \frac{8 \Omega^4}{\kappa^3} + \cdots
  \\
  {\cal C}&=&-\left(\gamma+\frac{4\Omega^2}{\kappa}\right)\left(1-\frac{4\Omega^2}{\kappa^2}\right)+\cdots
\end{eqnarray*}

Rewriting this equation equivalently in terms of the site occupied by the excitation, i.e. $\left|1\right>\equiv\left|eu\right>$ and $\left|2\right>\equiv\left|ue\right>$ yields
\begin{eqnarray*}
  \dot{\alpha}&=&\sum_{i=1}^2D_i\alpha D_i^\dag-\frac{1}{2}\left\{D_i^\dag D_i,\alpha \right\}+R_i\alpha R_i^\dag\\
  &-&\frac{1}{2}\left\{R_i^\dag R_i,\alpha \right\}+L_i\alpha L_i^\dag-\frac{1}{2}\left\{L_i^\dag L_i,\alpha \right\}.
\end{eqnarray*}
Here, the operator $D_i=\sqrt{\gamma_1}\left|i\right>\left<i\right|$ [with $\gamma_1=\frac{\gamma}{2}+\frac{4\Omega^2}{\kappa}-\frac{8\Omega^2}{\kappa^3}\left(3\Omega^2+\gamma^2\right)$] corresponds to the dephasing of the two states of the spin on site $i$, and the operators $R_i=\sqrt{{\cal D}/2}\left|i\right>\left<i+1\right|$ and $L_i=\sqrt{{\cal D}/2}\left|i\right>\left<i-1\right|$ to the hopping of the excitation to the right and left, respectively.
The time-evolution of the population of each site, i.e., the diagonal elements of the density matrix $\alpha^{i}\equiv\left<i\right|\alpha\left|i\right>$, is independent from the off-diagonal ones and read 
\begin{equation*}
  \partial_t\alpha^{i}=\frac{{\cal D}}{2}\left[\alpha^{i+1}+\alpha^{i-1}-2\alpha^{i}\right].
\end{equation*}
In the continuum limit ($\alpha^i\to\alpha(x)$), we obtain the following Fokker-Planck equation
\begin{equation*}
  \frac{\partial\alpha(x)}{\partial t}={\cal D}\frac{\partial^2\alpha(x)}{\partial x^2},
\end{equation*}
and we can identify the parameter ${\cal D}$ as a diffusion constant.

Note, that we have implicitly assumed that the case of two atoms is sufficient to explain the diffusion for a large number of atoms. In practice, there are further corrections to be included due to the finite size of the system. When performing a similar analysis for larger systems (up to $N=4$ sites), one finds that the actual form of the diffusion constant (up to fourth order in $\Omega$ and second in $\gamma$) is given by the Eq.\ (\ref{D}).

\begin{figure}[h]
\includegraphics[width=.8\columnwidth]{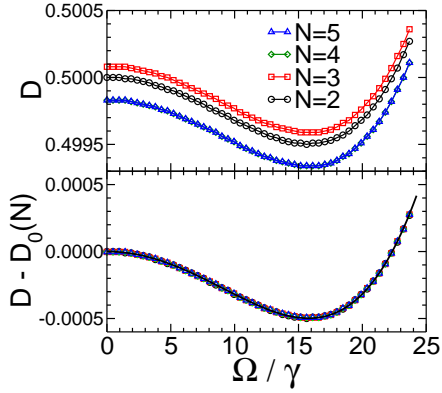}
\caption{The upper panel shows the diffusion constant from the numerically obtained master operator ${\mathbb W}_\alpha$ (\ref{Eq:alpha}) for lattices of size $N=2,3,4,5$.
The bottom panel shows collapse of the curves when the classical contribution is subtracted, $D-D_0(N)$, where $D_0$ is the value of the diffusion constant at $\Omega=0$. The values used are: $D_0(2)=\gamma/2$, $D_0(3)=\gamma/2+\gamma^2/12\kappa$ and $D_0(4)=D_0(5)=\gamma/2-\gamma^2/6\kappa$.}
\label{fig:DN}
\end{figure}

Finally, in order to test the validity of the analytic results, we have calculated numerically the master operator (\ref{Eq:alpha}) for small systems (up to $N=5$ sites) and obtained then the diffusion constant ($D$) and the branching ratio ($B$) as the element of ${\mathbb W}_\alpha$ connecting ${\boldsymbol e}{\boldsymbol u}{\boldsymbol u}\dots\rightarrow{\boldsymbol u}{\boldsymbol e}{\boldsymbol u}\dots$ and ${\boldsymbol e}{\boldsymbol u}{\boldsymbol u}\dots\rightarrow{\boldsymbol e}{\boldsymbol u}{\boldsymbol e}\dots$, respectively. The results of these simulations for the diffusion constant and their comparison with the analytic solution are shown in Figure \ref{fig:DN}.

\end{document}